\documentclass[prl,twocolumn,preprintnumbers,floatfix]{revtex4-1}

\usepackage{graphicx}
\usepackage{amsmath,amssymb,bm}
\usepackage{color}

\begin{document}

\title{
Higher-order Fermi-liquid corrections for an Anderson impurity away from half-filling
}

\author{Akira Oguri}
\affiliation{
Department of Physics, Osaka City University, Sumiyoshi-ku, 
Osaka 558-8585, Japan
}

\author{A.\ C.\ Hewson}
\affiliation{
Department of Mathematics, Imperial College London, London SW7 2AZ, 
United Kingdom
}

\date{\today}

\begin{abstract}
We study the higher-order Fermi-liquid relations of  Kondo systems 
for arbitrary impurity-electron fillings,  
 extending the many-body quantum theoretical approach of Yamada-Yosida. 
It includes partly  a microscopic clarification of the related achievements 
based on Nozi\`{e}res' phenomenological description: 
 Filippone,  Moca, von Delft, and Mora [Phys.\ Rev.\ B {\bf 95}, 165404 (2017) ].  
In our formulation,  the Fermi-liquid parameters such as 
the quasi-particle energy, damping, and transport coefficients 
are related to each other through the total vertex 
$\Gamma_{\sigma\sigma';\sigma'\sigma} (\omega, \omega'; \omega', \omega)$,  
which may be regarded as a generalized Landau quasi-particle interaction.  
We obtain exactly this function up to linear order with respect to 
the frequencies $\omega$ and $\omega'$ using the anti-symmetry and analytic properties. 
The coefficients acquire additional contributions of three-body fluctuations 
 away from half-filling through the non-linear susceptibilities. 
We also apply the formulation 
to non-equilibrium transport through a quantum dot,
and clarify how the zero-bias peak evolves in a magnetic field.

\end{abstract}


\maketitle

{\it Introduction.---}
Universal  low-energy behavior of interacting Fermi systems has been 
one of  the most fascinating properties in condensed matter physics.
Landau's Fermi liquid theory \cite{LandauFL,AGD,LandauLifshitz} 
phenomenologically explains  transport properties 
of electrons in a wide class of metals and normal liquid  $^3$He  
successfully \cite{LeggettHe3}, and may also be applied to exotic systems  
such as neutron stars and ultra-cold Fermi gases \cite{Bloch2012}. 
It starts with an expansion of the energy $E$ 
with respect to the deviation 
of  the momentum distribution function  $\delta n_{\bm{p}\sigma}^{}$ 
from the ground~state,
\begin{align}
E  \,=\, E_0 +
\sum_{\bm{p}\sigma} \varepsilon_{\bm{p}}^{} 
\delta n_{\bm{p}\sigma}^{} 
+ \frac{1}{2}
\sum_{\bm{p}\sigma \atop \bm{p}'\sigma'} 
f_{\bm{p}\, \sigma,\bm{p}' \sigma'}^{}\,
\delta n_{\bm{p}\sigma}^{} \delta n_{\bm{p}'\sigma'}^{} .  
\label{eq:Energy_functional}
\end{align}
The single quasi-particle energy $\varepsilon_{\bm{p}}^{}$ 
and the interaction between quasi-particles  $f_{\bm{p}\, \sigma,\bm{p}' \sigma'}^{}$ 
can microscopically be related to 
the four-point  vertex function,
 defined explicitly in the many-body quantum theory \cite{AGD,LandauLifshitz}.   
The field theoretic description has advantages over the phenomenologic approach:  
 the transport equations can be derived directly using  the Green's function  
without relying on empirical  assumptions nor 
the collision integral with the Boltzmann equation \cite{Eliashberg,EliashbergJETP15}.
For instance, through a microscopic consideration about 
the  anti-symmetry properties of the vertex function \cite{LandauLifshitz},  
sufficient conditions for the collective zero sound mode to exist 
have been derived \cite{MerminFL}.

Nozi\`{e}res extended the phenomenological Fermi-liquid description 
to Kondo systems \cite{NozieresFermiLiquid}, 
expanding the scattering phase shift $\delta$ with respect 
to a deviation of the occupation number of the impurity level in a way analogous to  
Eq.\ \eqref{eq:Energy_functional}.
Fully microscopic description was constructed 
by Yamada-Yosida, Shiba, 
and Yoshimori \cite{YamadaYosida2,YamadaYosida4,ShibaKorringa,Yoshimori}, 
and has also been extended to out-of-equilibrium quantum dots driven 
by a bias voltage $V$ \cite{Hershfield1,ao2001PRB}.
The two different types of descriptions complement each other 
and explain the universal behavior at temperatures $T$ 
 much lower than the Kondo energy scale $T_K$.
It is successful especially in the particle-hole symmetric case,   i.e.\ at half-filling,
where  the phase shift is locked at $\delta =\pi/2$ and     
the quadratic  $\omega^2$,  $T^2$ and $(eV)^2$ corrections  
emerge only through the quasi-particle {\it damping\/}.

Away from half-filling, however, the Kondo resonance peak deviates 
from the Fermi energy $\omega=0$, 
and as a consequence, the quadratic corrections emerge also through 
the real part of the self-energy  
due to the Coulomb interaction $U$ \cite{Yoshimori,HorvaticZlatic2}. 
It makes the problem difficult, 
and such corrections  have not been fully understood for a long time. 
Recently, there has been a significant breakthrough 
which shed light on this problem 
by extending Nozi\`{e}res' phenomenological description 
\cite{MoraMocaVonDelftZarand,FilipponeMocaVonDelftMora}.
Specifically, Filippone, Moca, von Delft and Mora (FMvDM) 
determined especially the quadratic coefficients of the self-energy 
away from half-filling 
\footnote{
Parameter correspondence:
$\alpha_{1\sigma}^{}/\pi =  \chi_{\sigma\sigma}^{}$,
$\phi_{1}^{}/\pi = - \chi_{\uparrow\downarrow}^{}$,
$2\,\alpha_{2\sigma}^{}/\pi =
 -\partial  \chi_{\sigma\sigma}^{} /\partial \epsilon_{d\sigma}^{}$, 
and 
$\phi_{2\sigma}^{}/\pi =  2\,  \partial  \chi_{\uparrow\downarrow}^{}/\partial \epsilon_{d\sigma}^{}$.}.

In this Letter, we provide a microscopic Fermi-liquid description 
for the non-equilibrium Anderson impurity \cite{AndersonModel} away from half-filling. 
One of the most pronounced merits of this formulation is that 
the real and imaginary parts of the transport coefficients are derived together from  
an explicit expression 
of the total vertex  $\Gamma_{\sigma\sigma';\sigma'\sigma} 
(\omega, \omega'; \omega', \omega)$ at low frequencies. 
It gives a clear answer to the long-standing problem. 
Specifically, an asymptotically exact expression is obtained, 
up to linear order in $\omega$ and  $\omega'$,
using the anti-symmetry and analytic properties with the Ward identities. 
The low-energy Fermi-liquid behavior is characterized by the expansion coefficients 
which are shown to be expressed in terms of 
the linear $\chi_{\sigma\sigma'}^{}$ and  non-linear 
$\chi_{\sigma_1\sigma_2\sigma_3}^{[3]}$  susceptibilities.

These susceptibilities can be calculated  
using  methods such as the numerical normalization group (NRG) \cite{KWW1} 
and the Bethe ansatz solution \cite{KawakamiOkiji,WiegmannTsvelick}.
We apply the microscopic formulation to  
non-equilibrium current $I$ through a quantum dot in the Kondo regime, 
and calculate the coefficients using the NRG. 
The result shows that the zero-bias peak of $dI/dV$ splits 
at a  magnetic field of the order of  $T_K$, 
and resolves a controversial issue about the splitting \cite{FilipponeMocaVonDelftMora}.
There are other numerical methods 
which work efficiently at different energy scales, 
such as the quantum Monte Carlo \cite{MillisWerner}, 
time-dependent NRG \cite{AndersTimeNRG} and 
density-matrix renormalization group \cite{Kirino}.  
Our approach has a numerical advantage at low energies 
as both the linear and non-linear susceptibilities can be deduced 
from the flow of energy eigenvalues near the fixed point of NRG 
\cite{HewsonOguriMeyer}.

The microscopic theory gives exact relations between different  response functions 
and has given  theoretical support  for the  universal scaling  observed in the  nonequilibrium currents  through quantum dots in the Kondo regime 
\cite{GrobisGoldhaber-Gordon,ScottNatelson}.
Furthermore, recent ultra-sensitive current noise measurements 
have successfully determined the Fermi-liquid parameters \cite{Ferrier2016} 
i.e.,  the Wilson ratio $R_W^{}$ and the renormalization factor of quasi-particles.
However, such comparisons so far have relied 
on the theoretical predictions at half-filling. 
The exact formula of transport coefficients,  
presented in Eqs.\ \eqref{eq:cT} and \eqref{eq:cV}, 
overcomes this restriction and can be applied to quantum dots 
for arbitrary electron fillings. 
Our formulation also has potential application for a wide class of Kondo systems  
such as dilute magnetic alloys  and quantum impurities 
with various kinds of internal degrees of freedom.

{\it Non-linear 3-body susceptibilities for impurity levels.---}
We consider the single Anderson impurity 
coupled to two noninteracting leads ($\lambda=L,R$); 
\begin{align}
\mathcal{H} =&  
 \sum_{\sigma}
 \epsilon_{d\sigma}^{}\, n_{d\sigma} 
 + U\,n_{d\uparrow}\,n_{d\downarrow} + 
\! \sum_{\lambda=L,R} \sum_{\sigma} 
\int_{-D}^D  \!\! d\epsilon\,  \epsilon\, 
 c^{\dagger}_{\epsilon \lambda \sigma} c_{\epsilon \lambda \sigma}^{}
\nonumber \\
& +   \sum_{\lambda=L,R} \sum_{\sigma}  v_{\lambda}^{}
 \left( \psi_{\lambda,\sigma}^\dag d_{\sigma}^{} + 
  d_{\sigma}^{\dag} \psi_{\lambda,\sigma}^{} \right) \;.
 \label{Hami_seri_part}
\end{align}
Here, 
 $d^{\dag}_{\sigma}$ creates an impurity electron 
with spin $\sigma$ 
 and $n_{d\sigma} = d^{\dag}_{\sigma} d^{}_{\sigma}$. 
Conduction electrons in each lead are normalized such that 
$
\{ c^{\phantom{\dagger}}_{\epsilon\lambda\sigma}, 
c^{\dagger}_{\epsilon'\lambda'\sigma'}
\} = \delta_{\lambda\lambda'} \,\delta_{\sigma\sigma'}   
\delta(\epsilon-\epsilon')$.  
In a magnetic field $h$, the impurity level is given by 
 $\epsilon_{d\sigma}^{} = \epsilon_{d}^{} - \sigma h$, 
 where  $\sigma = +1$ (-1) for $\uparrow$ ($\downarrow$) spin.
The hybridization $ v_{\lambda}^{}$ between
  $\psi^{}_{\lambda \sigma} \equiv  \int_{-D}^D d\epsilon \sqrt{\rho_c^{}} 
\, c^{\phantom{\dagger}}_{\epsilon\lambda \sigma}$ 
 and impurity electrons broadens the impurity level:     
 $\Delta \equiv \Gamma_L + \Gamma_R$ with 
 $\Gamma_{\lambda} = \pi \rho_c^{} v_{\lambda}^2$ 
and  $\rho_c^{}=1/(2D)$.  
We consider the parameter region, where  the  half band-width 
 $D$ is much greater than the other energy scales,    
$D \gg \max( U, \Delta, |\epsilon_{d\sigma}^{}|, |\omega|, T, eV)$.

We use  the  $T=0$  causal  impurity Green's function 
$G_{\sigma}^{}(\omega)$ and  self-energy $\Sigma_{\sigma}^{}(\omega)$ 
defined at  $eV=0$:
\begin{align}
& G_{\sigma}^{}(\omega) 
\,=\, \frac{1}{\omega -\epsilon_{d\sigma}
+i \Delta\,\mathrm{sgn} (\omega) -\Sigma_{\sigma}(\omega)} \;.
\end{align}
The phase shift 
$\cot \delta_{\sigma} \equiv {[\epsilon_{d\sigma} + \Sigma_\sigma(0)]}/{\Delta}$, 
or the density of states   $\rho_{d\sigma}^{} \equiv -
\mathrm{Im}\, G_{\sigma}^{}(0^+)/\pi$ 
at  $\omega=0$, 
 is a primary parameter which characterizes the 
Fermi-liquid ground state. 
The Friedel sum rule relates $\delta_{\sigma}$ to the occupation number  
which can also be given by the first derivative of the free energy 
  $\Omega \equiv - T \log \left[\mathrm{Tr}\,e^{-\mathcal{H}/T}\right]$,  
\begin{align}
&\langle n_{d\sigma} \rangle \,=\,
\frac{\partial \Omega}{\partial \epsilon_{d\sigma}} 
 \,\xrightarrow{\,T\to 0\,} \,  
\frac{\delta_{\sigma}}{\pi} 
\;.
\end{align}
The  leading Fermi-liquid corrections 
are determined by the static susceptibilities \cite{YamadaYosida2}, 
\begin{align}
\chi_{\sigma\sigma'}^{} \equiv \  
- \frac{\partial^2 \Omega}
{\partial \epsilon_{d\sigma'}\partial \epsilon_{d\sigma}} 
 \,=\,
- \,\frac{\partial \langle n_{d\sigma} \rangle }{\partial \epsilon_{d\sigma'}} 
\,\xrightarrow{\,T\to 0\,} \,  
 \rho_{d\sigma}^{} \, \widetilde{\chi}_{\sigma\sigma'}^{}.  
\label{eq:2body_susceptibility}
\end{align}
It can also be expressed as  $\chi_{\sigma\sigma'}^{} 
=  \int_0^{\frac{1}{T}}  \!d \tau 
\protect \langle \delta n_{d\sigma'}^{}(\tau) \delta  n_{d\sigma}^{} 
\protect \rangle$, and 
 $\widetilde{\chi}_{\sigma\sigma'}^{} 
\equiv 
\delta_{\sigma\sigma'} +
{\partial  \Sigma_{\sigma}(0)}/{\partial \epsilon_{d\sigma'}}$ 
is an enhancement factor similar to the Stoner factor.  
The usual spin and charge susceptibilities, 
 $\chi_s \equiv - \frac{1}{4} \frac{\partial^2 \Omega}{\partial h^2}$ and 
 $\chi_c\equiv  - \frac{\partial^2 \Omega}{\partial \epsilon_{d}^2} $, 
are given by  linear combinations of $\chi_{\sigma\sigma'}$ 
\footnote{ 
$ 
\chi_{s}  
= 
\frac{1}{4} \left(
\chi_{\uparrow\uparrow}^{} + \chi_{\downarrow\downarrow}^{} 
-  \chi_{\uparrow\downarrow}^{} - \chi_{\downarrow\uparrow}^{}
\right)$  and   
$\chi_{c}  
 =  
  \chi_{\uparrow\uparrow}^{} + \chi_{\downarrow\downarrow}^{} 
+  \chi_{\uparrow\downarrow}^{} +  \chi_{\downarrow\uparrow}^{}
$.  
 $\chi_{\sigma\sigma'}^{} = \chi_{\sigma'\sigma}^{}$ and  
 $\Omega$ is an even function of $h$.
}.
These susceptibilities also determine the characteristic energy scale 
 $4T^* \equiv 
1/\sqrt{\chi_{\uparrow\uparrow}^{}  \chi_{\downarrow\downarrow}^{}}$  
and  
the Wilson ratio  $R_W^{} \equiv  1 - 4T^* \chi_{\uparrow\downarrow}^{}$ 
which corresponds to a dimensionless quasi-particles 
interaction \cite{KWW1,HewsonRPT2001}.

Away from half-filling,  the third derivatives of the free energy 
also contribute to the next leading Fermi-liquid corrections,  
as we will show later   
\begin{align}
\chi_{\sigma_1\sigma_2\sigma_3}^{[3]} 
\equiv \  
- \,
\frac{\partial^3 \Omega }{\partial \epsilon_{d\sigma_1}^{}
\partial \epsilon_{d\sigma_2}^{}\partial \epsilon_{d\sigma_3}^{}} 
\,=\,  \frac{\partial \chi_{\sigma_2\sigma_3}}
{\partial \epsilon_{d\sigma_1}^{}} .
\end{align}
It can also be expressed as a static thee-point function 
of the impurity occupation 
 $\delta n_{d\sigma} \equiv n_{d\sigma} - \langle n_{d\sigma}  \rangle$, 
\begin{align}
\chi_{\sigma_1\sigma_2\sigma_3}^{[3]} \!  = 
- \!
\int_{0}^{\frac{1}{T}} \!\!\! d\tau_3 \!\! 
\int_{0}^{\frac{1}{T}} \!\!\! d\tau_2\, 
\langle T_\tau 
\delta n_{d\sigma_3} (\tau_3) \,
\delta n_{d\sigma_2} (\tau_2) \,
\delta n_{d\sigma_1}
\rangle .
\label{eq:canonical_correlation_3}
\end{align}

\begin{figure}[t]
 \leavevmode
\begin{minipage}{1\linewidth}
\includegraphics[width=0.5\linewidth]{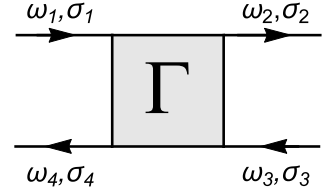}
\end{minipage}
 \caption{Total vertex  
$\Gamma_{\sigma_1\sigma_2;\sigma_3\sigma_4}^{}
(\omega_1, \omega_2; \omega_3, \omega_4)$  
satisfies the anti-symmetry property:   
Eq.\ \eqref{eq:vertex_antisymmetry} 
with $\omega_1+\omega_3=\omega_2+\omega_4$. 
}
 \label{fig:vertex}
\end{figure}

 {\it Higher-order Fermi-liquid corrections at $T=0$.---}
The Ward identity, 
which reflects the current conservation for each spin component  $\sigma$,  
plays a central role \cite{Yoshimori},
 \begin{align}
\frac{\partial \Sigma_{\sigma}(\omega)}{\partial  \omega}  
\, \delta_{\sigma\sigma'} 
+\frac{\partial \Sigma_{\sigma}(\omega)}{\partial \epsilon_{d\sigma'}}
= \, 
-
\Gamma_{\sigma \sigma';\sigma' \sigma}(\omega , 0; 0 , \omega) 
\rho_{d\sigma'}^{} .
\label{eq:YYY_causal}
\end{align}
Here, the total vertex  
  $\Gamma_{\sigma_1\sigma_2;\sigma_3\sigma_4}
(\omega_1, \omega_2; \omega_3, \omega_4)$  
includes all contributions of multiple scattering,  
and Fig.\ \ref{fig:vertex} shows the assignment of  arguments.   
The anti-symmetry properties of the total vertex 
also impose strong restrictions on the low-energy behavior    
as a consequence of the exclusion principle  \cite{AGD,LandauLifshitz,MerminFL,Rohringer},  
\begin{align}
&
\Gamma_{\sigma_1\sigma_2;\sigma_3\sigma_4}
(\omega_1, \omega_2; \omega_3, \omega_4) 
=   
-\Gamma_{\sigma_3\sigma_2;\sigma_1\sigma_4}
(\omega_3, \omega_2; \omega_1, \omega_4) 
\nonumber 
\\
&
\!\! 
=   
\Gamma_{\sigma_3\sigma_4;\sigma_1\sigma_2}
(\omega_3, \omega_4; \omega_1, \omega_2) 
= 
-\Gamma_{\sigma_1\sigma_4;\sigma_3\sigma_2} 
(\omega_1, \omega_4; \omega_3, \omega_2)  . 
\label{eq:vertex_antisymmetry}
\end{align}
For instance, at zero frequencies  the parallel-spin component vanishes   
 $\Gamma_{\sigma\sigma;\sigma\sigma}(0 , 0; 0 , 0)=0$, 
and the leading Fermi-liquid relations 
\footnote{
$ \protect \widetilde{\chi}_{\sigma\sigma'}  
+
\Gamma_{\sigma\sigma';\sigma'\sigma}(0 , 0; 0 , 0) 
 \, \rho_{d\sigma'} \,=\,
\bigl( 1-
\frac{\partial \Sigma_\sigma(\omega)}{\partial \omega}
|_{\omega=0}^{} \bigr)\,\delta_{\sigma \sigma'}$.} follow  from Eq.\ \eqref{eq:YYY_causal}.

Another important clue is the analytic property. 
Non-analytic part of the vertex function 
 is accompanied by the \lq\lq$\mathrm{sgn}$'' functions and is pure imaginary,  
while  the analytic part is real.
Thus,  the low-frequency expansion of 
the real part of  $\Gamma_{\sigma\sigma;\sigma\sigma}
(\omega_1, \omega_2; \omega_3, \omega_4)$  
starts with  a homogeneous polynomial of degree one. 
However, such a homogeneous polynomial of linear form cannot 
satisfy the anti-symmetry property  Eq.\ \eqref{eq:vertex_antisymmetry} 
provided $\omega_1+\omega_3=\omega_2+\omega_4$. 
Therefore,  the parallel-spin component does not have an analytic part of linear order. 
Thus,  for $\omega_2=\omega_3=0$,  
\begin{align}
&\left. 
\frac{\partial}{\partial \omega}
\mathrm{Re}\, 
\Gamma_{\sigma\sigma;\sigma\sigma}(\omega , 0; 0, \omega) 
\,\rho_{d\sigma}^{}\right|_{\omega\to 0}^{}  =\, 0 \;.
\label{eq:vert_UU_real_w}
\end{align}
To our knowledges, this property has not explicitly been recognized so far.  
We have also calculated the skeleton diagrams for 
$\Gamma_{\sigma\sigma;\sigma\sigma}(\omega , 0; 0, \omega)$ 
 up to order $U^4$ and have confirmed 
Eq.\ \eqref{eq:vert_UU_real_w}  perturbatively  
\footnote{Details will be given elsewhere:   
A.\ Oguri and A.\ C.\ Hewson, 
Phys.\ Rev.\ B {\bf 97}, 035435 (2018);
 Phys.\ Rev.\ B {\bf 97}, 045406 (2018). 
}.
In the linear order, the non-analytic part shows 
the  $|\omega|$ dependence \cite{ShibaKorringa} 
with a coefficient determined by Yamada-Yosida \cite{YamadaYosida4}:
\begin{align}
\Gamma_{\sigma\sigma;\sigma\sigma}(\omega , 0; 0, \omega) 
\,\rho_{d\sigma}^2 
\, =  \,  
 i \pi \,\chi_{\uparrow\downarrow}^2
\, \omega \, \mathrm{sgn} (\omega) 
+ O(\omega^2) . 
\label{eq:GammaUU_general_causal}
\end{align}
A series of higher-order Fermi-liquid relations follow 
from this property of the total vertex for parallel spins.

We obtain an identity between the double derivatives of the real part of the self-energy  
using Eqs.\ \eqref{eq:YYY_causal} and \eqref{eq:vert_UU_real_w}, 
\begin{align}
&
\mathrm{Re} \left. 
\frac{\partial^2 \Sigma_{\sigma}(\omega)
}{\partial \omega^2} \right|_{\omega \to 0}^{} 
= \,\frac{\partial^2  \Sigma_{\sigma}(0)}
{\partial \epsilon_{d\sigma}^2}
\;. 
\label{eq:self_real_w2}
\end{align}
 Note that 
${\partial^2  \Sigma_{\sigma}(0)}/{\partial \epsilon_{d\sigma}^2} \equiv 
 {\partial \widetilde{\chi}_{\sigma\sigma}}/{\partial \epsilon_{d\sigma}}$ 
by definition,  and   Eq.\ \eqref{eq:self_real_w2} agrees with FMvDM's result  
given in Eq.\ (B8b) of Ref.\ \onlinecite{FilipponeMocaVonDelftMora}.  
Furthermore, 
using Eqs.\  \eqref{eq:YYY_causal} and \eqref{eq:self_real_w2},
the total vertex for anti-parallel spins 
can  be calculated exactly up to terms of order  $\omega^2$, 
\begin{align}
 &
\!\!
\Gamma_{\sigma, -\sigma;-\sigma, \sigma}(\omega, 0; 0 ,\omega) 
\,\rho_{d\sigma}^{}\rho_{d,-\sigma}^{}
\ = \ 
\,-\chi_{\uparrow\downarrow} + 
\rho_{d\sigma}
\frac{\partial \widetilde{\chi}_{\sigma,-\sigma}}
{\partial \epsilon_{d\sigma}} \, \omega  
\nonumber \\
& 
\!\!
 +\frac{\rho_{d\sigma}}{2} \,
\frac{\partial}{\partial \epsilon_{d,-\sigma}} 
\left[
- 
\frac{\partial \widetilde{\chi}_{\sigma\sigma}}{\partial \epsilon_{d\sigma}} 
 + i \pi\, \frac{\chi_{\uparrow\downarrow}^2}{\rho_{d\sigma}^{}} 
 \, \mathrm{sgn}(\omega) 
\right]  \omega^2  + \cdots .
\label{eq:GammaUD_general_causal}
\end{align}
Note that the $\omega$-linear contribution  is real and analytic.

We see in Eqs.\ \eqref{eq:self_real_w2} and \eqref{eq:GammaUD_general_causal} 
that expansion coefficients depends on 
${\partial \widetilde{\chi}_{\sigma\sigma'}}/{\partial \epsilon_{d\sigma''}}$ 
which includes contributions from three-body fluctuations 
$\chi_{\sigma\sigma'\sigma''}^{[3]}$.  
The three-body correlations vanish in the particle-hole symmetric case  
since  the spin (charge) susceptibility takes a maximum (minimum): 
${\partial \chi_s}/{\partial \epsilon_d^{}} =0$  and 
${\partial \chi_c}/{\partial \epsilon_d^{}} =0$ 
at $\xi_d \equiv  \epsilon_d +U/2=0$ and $h=0$. 
We also find  that the $\omega^2$ term of  Eq.\ \eqref{eq:GammaUD_general_causal} 
involves  four-body fluctuations in the real part through 
${\partial^2 \widetilde{\chi}_{\sigma\sigma}}
/{\partial \epsilon_{d\sigma}\partial \epsilon_{d,-\sigma}}$  
 which remains finite  even in the particle-hole symmetric case.
The four-body fluctuations 
will also contribute to higher-order terms of the parallel-spin vertex.

We have also calculated the total vertex for two independent frequencies 
up to linear order in  $\omega$ and $\omega'$:  
\begin{align}
\Gamma_{\sigma\sigma;\sigma\sigma}(\omega , \omega'; \omega', \omega) 
\,\rho_{d\sigma}^{2}
\, =  \, 
 i \pi \,
\chi_{\uparrow\downarrow}^2
\,\bigl|\omega-\omega' \bigr| 
+ \cdots ,
\label{eq:GammaUU_general_omega_dash}
\end{align}
\begin{align}
&
\!\!\!\!\!\!\!\!\!\!\!\!\!\!\!\!\!
\Gamma_{\sigma, -\sigma;-\sigma, \sigma}(\omega, \omega'; \omega' ,\omega) 
\,\rho_{d\sigma}^{}\rho_{d,-\sigma}^{}
\nonumber \\ 
 = &  \ 
-\chi_{\uparrow\downarrow} + 
\rho_{d\sigma}^{}
\frac{\partial \widetilde{\chi}_{\sigma,-\sigma}}
{\partial \epsilon_{d\sigma}^{}} \, \omega  
+ 
\rho_{d,-\sigma}^{}
\frac{\partial \widetilde{\chi}_{-\sigma,\sigma}}
{\partial \epsilon_{d,-\sigma}^{}} \, \omega'   
\nonumber \\
&  + i \pi \,\chi_{\uparrow\downarrow}^2
\Bigl(
\,\bigl|  \omega - \omega'\bigr| 
-
\,\bigl| \omega + \omega' \bigr| 
\Bigr)
+ \cdots
.
\label{eq:GammaUD_general_omega_dash}
\end{align}
The analytic real part can be deduced   
from Eqs.\ \eqref{eq:GammaUU_general_causal} 
and \eqref{eq:GammaUD_general_causal} 
using the anti-symmetry properties Eq.\ \eqref{eq:vertex_antisymmetry}. 
The non-analytic part has been obtained through  
 an additional consideration about the singular Green's-function products 
 \cite{EliashbergJETP15,Yoshimori,ao2001PRB}.
Specifically, the $|\omega-\omega'|$ and $|\omega+\omega'|$ contributions 
emerge from the intermediate particle-hole and 
particle-particle pair excitations, respectively. 
We note that the total vertex, Eqs.\ 
\eqref{eq:GammaUU_general_omega_dash} and 
\eqref{eq:GammaUD_general_omega_dash},  
 can be regarded as a quantum-impurity analogue of 
 Landau's phemomenological interaction $f_{\bm{p}\sigma,\bm{p}' \sigma'}^{}$,  
 and can also be compared  
with Nozi\`{e}res' function $\phi_{\sigma\sigma'}(\varepsilon,\varepsilon')$ 
 \cite{LandauFL,NozieresFermiLiquid}. 
One of the advantages of the microscopic formulation 
to the phenomenological descriptions is that the real and imaginary parts, 
which contribute to the energy-shift  and damping 
of quasi-particles,  are described  in a unified way 
with clearly defined correlation functions.

 {\it The  $T^2$ and $(eV)^2$ self-energy corrections.---}
The $T^2$ correction of the retarded self-energy 
$\Sigma_{\sigma}^r(\omega,T) $ can be deduced from the derivative of 
$\Gamma_{\sigma\sigma';\sigma' \sigma}(\omega, \omega'; \omega' ,\omega)$ 
with respect to $\omega'$  using the formula \cite{YamadaYosida4,AGD,Note4},  
\begin{align}
& 
\Sigma_{\sigma}^r(0,T) 
- \Sigma_{\sigma}^r(0,0)
 = \, 
\frac{(\pi  T)^2}{6}
\lim_{\omega \to 0^+} 
\Psi_{\sigma}^{}(\omega)   +  
\cdots
\label{eq:Psi_result_T2}
\\
&\Psi_{\sigma}^{}(\omega) 
\equiv  
 \lim_{\omega' \to 0}
\frac{\partial}{\partial \omega'} 
 \sum_{\sigma'}
\Gamma_{\sigma \sigma';\sigma' \sigma}(\omega, \omega'; \omega', \omega) 
\rho_{d\sigma'}^{}(\omega') .
\label{eq:Psi_T0}
 \end{align}
Substituting Eqs.\ \eqref{eq:GammaUU_general_omega_dash} 
and \eqref{eq:GammaUD_general_omega_dash} into Eq.\ \eqref{eq:Psi_T0}   
 \footnote{Here,
 $\rho_{d\sigma}^{}(\omega) \equiv 
 (-1/\pi)\,\mathrm{Im}\,G_{\sigma}^{r}(\omega)$,  in the retarded form.
 }, we obtain 
 \begin{align}
 \lim_{\omega \to 0} 
\Psi_{\sigma}^{}(\omega)  
\, =&  \ 
 \frac{1}{\rho_{d\sigma}^{}} 
\frac{\partial \chi_{\uparrow\downarrow}}{\partial \epsilon_{d,-\sigma}} 
\, - i \,3\,\pi \, \frac{\chi_{\uparrow\downarrow}^2}{\rho_{d\sigma}^{}}
\, \mbox{sgn}(\omega) 
\;.
\label{eq:Psi_result_+}
\end{align}
Here, 
the real part, ${\partial \chi_{\uparrow\downarrow}}/{\partial \epsilon_{d,-\sigma}}$, 
emerges from the analytic part of the total vertex for anti-parallel spins.  
%

 In a previous work, we have diagrammatically shown 
that the low-bias  $(eV)^2$ self-energy 
can be calculated taking a variational derivative of the equilibrium self-energy 
with respect to the internal Green's functions \cite{ao2001PRB,ao2005jpsj}.  
Revisiting the details of the calculation, 
we find exactly the same quantum-mechanical intermediate states, 
which consequently lead to Eq.\ \eqref{eq:Psi_result_+}, 
determine both the  $(eV)^2$  and $T^2$ corrections 
\footnote{
Eq.\ \eqref{eq:Psi_T0} of this paper is 
identical to Eq.\ (14) of Ref.\ \onlinecite{ao2001PRB}  i.e., 
 $\Psi_{\sigma}^{}(\omega) =  
\protect \widehat{D}^2  \Sigma_{\sigma}^{}(\omega)$ 
with  $\protect \widehat{D}^2$ defined in Ref.\ \onlinecite{ao2001PRB} 
}.  
The relation between these two corrections  
has been first pointed out by FMvDM using  Nozi\`{e}res' description 
\cite{FilipponeMocaVonDelftMora}. 
Our result provides a complete proof for this observation.

Using the above results, low-energy behavior of 
the retarded self-energy $\Sigma_{\sigma}^r(\omega,T,eV)$  
 is exactly determined up to terms of order $\omega^2$, $T^2$, and $(eV)^2$. 
To be specific, 
the bias voltage $eV  \equiv \mu_L-\mu_R$ is applied 
through the chemical potentials of the left and right leads,   
 $\mu_L= \alpha_L eV$ and  $\mu_R= -\alpha_R\, eV$,    
 with additional parameters satisfying $\alpha_L + \alpha_R =1$. 
Thus, the self-energy generally depends not only on $eV$ but also    
 $\alpha \equiv 
 (\alpha_L \Gamma_L - \alpha_R \Gamma_R)/(\Gamma_L+ \Gamma_R)$ 
\cite{ao2001PRB}. 
The  asymptotically exact  imaginary and real parts of the retarded self-energy 
are given by 
 \begin{align}
&
\!\!\!
\mbox{Im}\, \Sigma_{\sigma}^r(\omega,T,eV) 
  = 
 - \frac{\pi}{2}\,   
\frac{\chi_{\uparrow\downarrow}^2}{\rho_{d\sigma}^{}}
 \nonumber  \\
& 
\!\!\!\!
 \times   \biggl[
    \left(\,\omega -   \alpha\, eV\,  \right)^2 
+ \frac{ 3\,\Gamma_L \Gamma_R}{\left( \Gamma_L + \Gamma_R \right)^2} \,(eV)^2 
 +(\pi T)^2  
    \biggr]  + \cdots .
\label{eq:self_imaginary}
\end{align}
\begin{align}
 &
\epsilon_{d\sigma} +  \mathrm{Re}\, \Sigma_{\sigma}^r(\omega,T,eV) 
 \  = \ 
 \Delta\, \cot \delta_{\sigma}
 + \bigl( 1-\widetilde{\chi}_{\sigma\sigma} \bigr)\, \omega 
 \nonumber \\ 
  & 
   + \frac{1}{2} \frac{\partial \widetilde{\chi}_{\sigma\sigma}}
 {\partial \epsilon_{d\sigma}}  \omega^2 
  +  
 \frac{1}{6\rho_{d\sigma}} 
 \frac{\partial \chi_{\uparrow\downarrow}}{\partial \epsilon_{d,-\sigma}} 
 \left[
 \frac{3\Gamma_L \Gamma_R}{\left( \Gamma_L + \Gamma_R \right)^2} 
  (eV)^2 
 +
  \left( \pi T\right)^2 
 \right] 
 \nonumber \\ 
  & 
 - \widetilde{\chi}_{\sigma,-\sigma}^{}  \alpha\,  eV 
 + 
  \frac{\partial \widetilde{\chi}_{\sigma,-\sigma}}{\partial \epsilon_{d\sigma}} 
  \alpha\,  eV \omega
  +  \frac{1}{2}
 \frac{\partial\widetilde{\chi}_{\sigma,-\sigma}}{\partial\epsilon_{d,-\sigma}}
 \,\alpha^2 (eV)^2 
+ \cdots  .
 \label{eq:self_real_ev_mag}
 \end{align}
We note that Eq.\ \eqref{eq:self_real_ev_mag}  is consistent 
with the previous result of ours \cite{ao2001PRB},   
derived for general electron-fillings 
without the knowledge of Eq.\ \eqref{eq:self_real_w2}
\footnote{
Substituting 
the  $\omega^2$ real part given in Eq.\ \eqref{eq:self_real_w2} 
into  Eq.\ (19) of Ref.\ \onlinecite{ao2001PRB},  
we obtain the expression which agrees with 
Eq.\ \eqref{eq:self_real_ev_mag} at $h\to 0$. 
}. 
Equation \eqref{eq:self_real_ev_mag} 
is a generalized formula of the real part, 
which also extends  FMvDM's result \cite{FilipponeMocaVonDelftMora} 
to asymmetric junctions $\alpha \neq 0$ \cite{Note4}.

{\it  Non-equilibrium magneto transport.---}
We next consider the current flowing through the Anderson impurity  $I$
\footnote{$I =
\frac{e}{2\pi\hbar} \sum_\sigma \int \! 
d\omega \, 
\frac{4\Gamma_R\Gamma_L}{\Gamma_R+\Gamma_L} 
\left[ f_L^{}(\omega) - f_R^{}(\omega)\right] 
\,\pi \rho_{d\sigma}^{}(\omega)$, where  
$f_\lambda(\omega)\equiv f(\omega-\mu_{\lambda})$ and  
 $f(\omega)=[e^{\omega/T}+1]^{-1}$. See also 
 \cite{Note5}}, 
 using the Meir-Wingreen formula \cite{MeirWingreen} 
with Eqs.\ \eqref{eq:self_imaginary} and  \eqref{eq:self_real_ev_mag}. 
Specifically, we examine  a symmetric junction with         
 $\Gamma_L=\Gamma_R$ and  $\mu_L=-\mu_R = eV/2$,  
for which the conductance can be expressed in the form
\begin{align}
&
\!\!\!
\frac{dI}{dV}   
 =  \,  
 \frac{e^2}{2\pi \hbar}  \sum_{\sigma} 
\left[  
 \sin^2 \delta_{\sigma}  
-  c_{T,\sigma}^{}
      \left(\pi T  \right)^2
-  c_{V,\sigma}^{}
         \left(eV\right)^2 
 \right] \! , 
\end{align}
\begin{align}
&
\!\!
c_{T,\sigma}^{} 
=  \  
\frac{\pi^2}{3} 
\biggl[\,
-
 \cos 2 \delta_{\sigma}
\left(
\chi_{\sigma\sigma}^2
+ 
2
\chi_{\uparrow\downarrow}^2 
\right)
\,
\nonumber \\
&  \qquad \qquad \quad
+
\frac{\sin 2\delta_{\sigma}}{2\pi}\,
\left(
\frac{\partial \chi_{\sigma\sigma}}{\partial \epsilon_{d}^{}}
+ \sigma \frac{\partial \chi_{\uparrow\downarrow}}{\partial h} 
\right)
\,\biggr]
, 
\label{eq:cT}
\\
& 
\!\!
c_{V,\sigma}^{}   
 =  \   
\frac{\pi^2}{4}
\biggl[
\,
-
\cos 2 \delta_{\sigma} 
\left( \chi_{\sigma\sigma}^2 +  5\,\chi_{\uparrow\downarrow}^2
\right)
\nonumber \\
&  \qquad \qquad    
+ 
\frac{\sin 2\delta_{\sigma}}{2\pi}
\left(
\frac{\partial \chi_{\sigma\sigma}}{\partial \epsilon_{d}^{}} 
+
\frac{\partial \chi_{\uparrow\downarrow}}{\partial \epsilon_{d}^{}} 
+\sigma\, 2  
\frac{\partial \chi_{\uparrow\downarrow}}{\partial h} 
\right) \,
\biggr] .
\label{eq:cV}
\end{align}
Here, contributions of the three-body fluctuations 
enter through the derivatives of susceptibilities 
with respect to $\epsilon_d^{}$ or $h$,  
which are accompanied by the factor  $\sin 2 \delta_{\sigma}$.   
For the magneto conductance in the Kondo regime, 
there is a controversial issue \cite{FilipponeMocaVonDelftMora}:
whether or not the zero-bias peak of $dI/dV$ 
splits at a magnetic field of the order of the Kondo energy scale $T_K$. 
We demonstrate in the following that 
calculations with the exact conductance formula, 
 Eqs.\ \eqref{eq:cT} and \eqref{eq:cV},  
resolve the problem \cite{Note4}.

\begin{figure}[t]
 \leavevmode

\begin{minipage}{1\linewidth}
\includegraphics[width=0.72\linewidth]{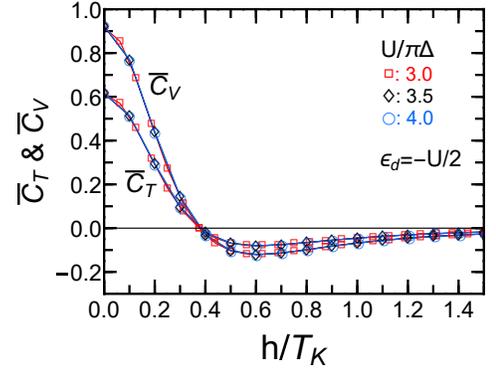}
\end{minipage}
 \caption{
(Color online) Magnetic field dependence of the coefficients  
 $\overline{C}_{T}^{} = T_K^2 \sum_{\sigma} c_{T,\sigma}^{}/2 $ and 
 $\overline{C}_{V}^{}  = T_K^2 \sum_{\sigma} c_{V,\sigma}^{}/2$ 
 at  $\epsilon_d^{}=-U/2$. 
 $T_K =  z_0 \pi \Delta/4$ is  defined at  $h=0$ with 
the renormalization factor  $z_0 \simeq 0.08$,  $0.05$, and  $0.03$  
for $U/\pi \Delta = 3.0$, $3.5$, and $4.0$, respectively.
At $h=0$ the coefficients approach 
$\overline{C}_{T}^{} \to \pi^2/16$  and $\overline{C}_{V}^{} \to 3\pi^2/32$  
in the  $U\to \infty$ limit.
} 
 \label{fig:field_dependence_half_filling}
\end{figure}

We have calculated the phase shift  $\delta_{\sigma}^{}$ and
the enhancement factor  $\widetilde{\chi}_{\sigma\sigma'}$ 
as functions of $h$ at $\epsilon_d = -U/2$ using the NRG 
\cite{HewsonBauerKoller,Hewson2011NATO}.
The dimensionless coefficients  
 $\overline{C}_{T}^{} = T_K^2 \sum_{\sigma} c_{T,\sigma}^{}/2 $ and 
 $\overline{C}_{V}^{}  = T_K^2 \sum_{\sigma} c_{V,\sigma}^{}/2$ 
have been determined substituting the NRG results into 
 Eqs.\ \eqref{eq:cT} and \eqref{eq:cV}.  
The result is shown in Fig.\  \ref{fig:field_dependence_half_filling} 
as a function of  $h/T_K$,  using  $T_K=1/4\chi_{\uparrow\uparrow}$ 
defined at $h=0$ for each case of $U/\pi \Delta$ ($=3.0,3.5,4.0 $) 
\footnote{
We used  $\Lambda=2.0$ for 
the discretization parameter 
and keep $3600$ states per iteration \cite{KWW1}
}.
We see that both  $\overline{C}_{T}^{}$ and   $\overline{C}_{V}^{}$   
show the universal Kondo behavior. 
This is consistent with the behavior of the Wilson ratio which 
 is almost saturated to the strong-coupling value $R_W \simeq 2$ 
for $U/\pi \Delta \gtrsim 3$ \cite{HewsonBauerKoller}. 
Furthermore, $\overline{C}_{T}^{}$ and  $\overline{C}_{V}^{}$  change 
sign at $h$  of order $T_K$:   
at very close magnetic-field values  $h \simeq 0.38 T_K$.  
This means that the zero-bias peak does split for  $h\gtrsim 0.38 T_K$   
because  $dI/dV$  increases from the zero-bias value as  $eV$ increases.
These observations  are consistent with the previous second-order 
renormalized perturbation result \cite{HewsonBauerOguri}.

{\it Conclusion.---} 
We have provided a many-body quantum theoretical description of  
 the Fermi-liquid state in the particle-hole asymmetric case. 
The Fermi-liquid corrections away from half-filling are  
characterized by additional contributions of the 
 three-body fluctuations which enter through 
the non-linear response function   $\chi_{\sigma_1\sigma_2\sigma_3}^{[3]}$.
The asymptotically exact expression of the total vertex  
 $\Gamma_{\sigma\sigma';\sigma'\sigma}
(\omega, \omega'; \omega', \omega)$ 
describes low-energy properties in a unified way: 
this function and its derivatives with respect to $\omega$ or $\omega'$ 
determine the quasi-particle interaction,  
energy shift, damping, and transport coefficients can be 
generated systematically up to order $\omega^2$, $T^2$, and $(eV)^2$,  
with the Ward identities given in  Eqs.\ \eqref{eq:YYY_causal} and \eqref{eq:Psi_T0}.
Furthermore, the non-equilibrium self-energy Eq.\  \eqref{eq:self_real_ev_mag} 
is applicable to the asymmetric tunneling couplings, 
and has potential application for real quantum dots 
\cite{GrobisGoldhaber-Gordon,ScottNatelson,Ferrier2016}.
We have also demonstrated an application to the 
non-linear magnetoconductance through a quantum dot in the Kondo regime,  
and have shown that  the zero-bias peak of $dI/dV$ 
 splits naturally at a magnetic finite field of order $T_K$.
Our description can be extended, and may be  used,  
to explore a wide class of Kondo systems 
and more general quantum impurities.

 \bigskip

We wish to  thank J.\ Bauer and R.\ Sakano 
for valuable discussions, and C.\ Mora and J.\ von Delft for 
sending us Ref.\ \onlinecite{FilipponeMocaVonDelftMora} prior to publication.
This work was supported by JSPS KAKENHI (No.\ 26400319) and 
  a Grant-in-Aid for Scientific Research (S) (No.\ 26220711).


%

\end{document}